\documentclass[12pt,letter]{article}%
\pdfoutput=1

\usepackage[T1]{fontenc}
\usepackage[utf8]{inputenc}

\DeclareUnicodeCharacter{00DF}{\ss}

\usepackage{amsmath}
\usepackage{amssymb}
\usepackage{amsthm}
\usepackage[ 
    ]{hyperref}
\usepackage{mathtools}
\usepackage{graphicx}
\usepackage{booktabs}
\usepackage{caption}
\usepackage{subfig}
\usepackage{slashed}
\usepackage{listings}
\usepackage{slashed}
\usepackage{rotating}
\usepackage{feynmp}
\usepackage{subfig}
\usepackage{geometry}
\usepackage{multicol}
\usepackage{dsfont}
\usepackage[nottoc]{tocbibind}
\usepackage{multicol}
\usepackage{authblk}
\usepackage{cite}
\usepackage{color}

\numberwithin{equation}{section} 
\setcaptionmargin{1cm}

\begin{document}

\vspace{-0.2cm}
\rightline{DESY-15-242}
\vspace{1.2cm}
\begin{center}


\LARGE{\bf Axion Monodromy \\ and the Weak Gravity Conjecture} 
\\[13mm]
  \large{Arthur Hebecker,$^{1}$ Fabrizio Rompineve$^{1}$, and Alexander Westphal$^{2}$ }
  \\[13mm]
  \small{
 $^1$\,\emph{Institute for Theoretical Physics, University of Heidelberg, Philosophenweg 19, 69120 Heidelberg, Germany}\quad \quad \\
$^2$\,\emph{DESY, Theory Group, Notkestra{\ss}e 85, D-22603 Hamburg, Germany} 
\\[8mm]}
\small{\bf Abstract} \\[7mm]
\end{center}
\begin{center}
\begin{minipage}[h]{15.22cm}
Axions with broken discrete shift symmetry (axion monodromy) have recently played a central role both in the discussion of inflation and the `relaxion' approach to the hierarchy problem. We suggest a very minimalist way to constrain such models by the weak gravity conjecture for domain walls: While the electric side of the conjecture is always satisfied if the cosine-oscillations of the axion potential are sufficiently small, the magnetic side imposes a cutoff, $\Lambda^3 \sim m f M_{pl}$, independent of the height of these `wiggles'. We compare our approach with the recent related proposal by Ibanez, Montero, Uranga and Valenzuela. We also discuss the non-trivial question which version, if any, of the weak gravity conjecture for domain walls should hold. In particular, we show that string compactifications with branes of different dimensions wrapped on different cycles lead to a `geometric weak gravity conjecture' relating volumes of cycles, norms of corresponding forms and the volume of the compact space. Imposing this `geometric conjecture', e.g.~on the basis of the more widely accepted weak gravity conjecture for particles, provides at least some support for the (electric and magnetic) conjecture for domain walls.

\end{minipage}
\end{center}
\vspace*{10ex}
December 11, 2015

\newpage

\begin{section}{Introduction}

The fog surrounding Large Field Inflation is rapidly dissolving. On the one hand, if experiments are going to detect primordial gravitational waves in the very near future, then we will know that trans-Planckian field displacements are required. On the other hand, ongoing theoretical effort may rule out large field inflation in effective field theory coupled to gravity before the observational verdict. In particular, recent constraints focus on models where the inflaton is an axion with a super-Planckian decay constant.

Axions with large periodicities also occur in a different setting: they are a crucial ingredient of the proposed \emph{relaxion} solution to the hierarchy problem \cite{Graham:2015cka} (see also \cite{Espinosa:2015eda, Hardy:2015laa, Patil:2015oxa, Jaeckel:2015txa, Gupta:2015uea, Batell:2015fma, Choi:2015fiu, Kaplan:2015fuy}).
 
However, trans-Planckian values of decay constants are problematic \cite{Banks:2003sx}.\footnote{See \cite{Conlon:2016aea} for very recent work and \cite{Abel:2015rkm} for a recent discussion of the underlying shift symmetries beyond tree level.} Nevertheless, there exist several proposals to implement large-field axion inflation in effective field theory with sub-Planckian decay constants. They include decay constant alignment (KNP)\cite{Kim:2004rp} and N-flation \cite{Dimopoulos:2005ac} (see also \cite{Liddle:1998jc}). For a biased collection of recent implementations in string theory see \cite{Cicoli:2014sva, Palti:2014kza, Grimm:2014vva, Ibanez:2014kia, Kappl:2014lra, Ben-Dayan:2014zsa, Long:2014dta, Ben-Dayan:2014lca, Abe:2014xja, Ruehle:2015afa}. Another proposal, which is the focus of this paper, is axion monodromy \cite{Silverstein:2008sg} (see also \cite{Marchesano:2014mla, Blumenhagen:2014gta, Hebecker:2014eua, Blumenhagen:2014nba, Hebecker:2014kva, Ibanez:2014swa, Escobar:2015fda, Kobayashi:2015aaa, Escobar:2015ckf} for realisations in string theory).

Recently, important quantum gravity arguments have been used to constrain, and in some cases even rule out, many of these models (see \cite{ArkaniHamed:2006dz, Conlon:2012tz, Cheung:2014vva, Rudelius:2014wla, delaFuente:2014aca, Rudelius:2015xta, Montero:2015ofa, Brown:2015iha, Bachlechner:2015qja, Hebecker:2015rya,Brown:2015lia, Junghans:2015hba,Heidenreich:2015wga, Palti:2015xra, Heidenreich:2015nta,Kooner:2015rza, Andriot:2015aza, Kaloper:2015jcz}). A first criticism to large field displacements is based on gravitational instantons \cite{Montero:2015ofa}. A second one is based on the Weak Gravity Conjecture (WGC) \cite{ArkaniHamed:2006dz}. The latter has been successfully used to constrain models of $N$-flation and decay constant alignment \` {a} la KNP . The implications of the WGC for axion monodromy are less clear (see however \cite{Brown:2015iha}). A third criticism is based on entropy arguments (\hspace{-0.1ex}\cite{Conlon:2012tz}, see however \cite{Kaloper:2015jcz}). 
 
Very recently, developing an idea of \cite{Brown:2015iha}, the authors of \cite{Ibanez:2015fcv} have applied the WGC for domain walls to axion monodromy, albeit mainly in the different context of the relaxion proposal. Their analysis rests on interpreting the monodromy as being due to the gauging of the discrete shift symmetry of an axion by a 3-form potential \` {a} la Kaloper-Sorbo (KS) \cite{Kaloper:2008fb, Kaloper:2011jz} (see also \cite{Dvali:2005an}). The WGC for the original 3-form gauge theory says that this system comes with light domain walls which, in turn, threaten the slow-roll field evolution in the resulting monodromy model.

The WGC is a statement which connects low energy effective field theories and their UV completion (which we assume to be string theory). In the very same spirit, this paper follows two different, but related, directions: the first one is more phenomenological and relevant both to inflation and relaxation, while the second one is more conceptual and deals with string compactifications. The link is provided by the WGC.

Concerning the first direction, we advocate a different point of view on constraints coming from $4d$ membranes in models of axion monodromy (inflation and relaxation). In particular, we take a minimalist effective field theory perspective:  a generic realisation of monodromy is characterised by `wiggles' in the axion potential (see Fig. \ref{fig:potential1}, \ref{fig:potential2}). The latter define a four-form flux and associated domain walls. Those differ from the membranes analysed in \cite{Brown:2015iha, Ibanez:2015fcv}, as they do not arise from a gauging procedure: they originate purely from the oscillatory axion potential. In particular, their tension can be made lighter without spoiling slow-roll. In fact, the lower the `wiggles', the easier it is to slow-roll. Therefore the electric WGC does not constrain generic realisations of axion monodromy (inflation and relaxation). 

We then seek for constraints arising from the magnetic WGC. It is generically expected that axion monodromy models cannot allow for a parametrically large field range when correctly implemented in a setup of string compactifications. Our claim is that the magnetic WGC describes precisely such a limitation to the field range (see also \cite{Mazumdar:2014qea} for general constraints in setups of string compactifications). It does so by bounding the cutoff of the effective theory of an axion with periodicity $2\pi f$ and monodromy-induced potential $m^{2}\phi^{2}$: $\Lambda^{3}\lesssim mfM_{pl}$. Our point of view is relevant not only for models where monodromy is used to realise inflation, but also for relaxation models, where the low energy barriers are a fundamental ingredient of the mechanism. Applied to inflation, this condition allows in principle for large field displacements, but forbids models with a too small decay constant. Our constraint may be considered as a positive statement about the feasibility of Axion Monodromy Inflation. In addition, this drives limits on the amount of resonant non-Gaussianity~\cite{Chen:2008wn,Flauger:2009ab} from the `wiggles', and on the possibility of slow-roll eternal inflation.

We then justify our extension of the magnetic side of the WGC to domain walls.  In assuming that the electric WGC can be extended to any $p$-dimensional object in $d$ dimensions, we are motivated purely by the fact that string theory always fulfils the WGC. We argue that, adopting this point of view, the extension of the magnetic part is equally well motivated. In fact, we show that string theory fulfils the WGC precisely by lowering the cutoff of the $4d$ description, i.e. the KK scale, rather than by providing objects which are light enough. The electric side is therefore satisfied as a consequence of the magnetic WGC.

The second aim of this paper is to describe some insight concerning extensions of the WGC to generic $p$-dimensional object in $d$-dimensions. This has been already the focus of recent activity \cite{Brown:2015iha, Heidenreich:2015nta, Heidenreich:2015wga, Ibanez:2015fcv}. We make progress by showing that, in a setup of string compactifications, the WGC can be phrased as a purely geometric constraint. In particular, it translates into a requirement on the size and intersections of the $q$-cycles wrapped by the $p$-dimensional objects. Explicitly:
\begin{equation}
\frac{V_{X}^{1/2}|q^{\Sigma}|}{V^{\Sigma}}\geq A_{d},
\end{equation} 
where $V_{X}$ is the volume of the compactification manifold, $V^{\Sigma}$ is the volume of the $q$-dimensional cycle $\Sigma$, $|q^{\Sigma}|$ is the norm of the harmonic form related to $\Sigma$ using the metric $X$, and $A_{d}$ is a $O(1)$ number given below.
Once assumed for one particular configuration (e.g. one leading to $4d$ particles), this constraint is valid for any other $s$-dimensional object, with $s\neq p$, wrapped on the same cycles of the same CY. Therefore, our strategy shows that string dualities are not needed to generalise the WGC: one can separately constrain theories with different brane configurations compactified on the same CY. In this sense our approach improves on the existing literature.

This paper is structured as follows. Sect.~\ref{sec:pheno} is devoted to phenomenological considerations. In particular, in Sect.~\ref{sub:domainwalls1} we describe the presence of domain walls in $4d$ effective field theory models of axion monodromy and deduce the constraints coming from the electric WGC. In Sect.~\ref{sec:domainwalls2} we assume the magnetic WGC for domain walls and extract the consequences for Axion Monodromy Inflation. In Sect.~\ref{sub:motivation} we motivate the extension of the magnetic WGC to domain walls and in Sect.~\ref{sub:ks} we comment on the relation to KS membranes. Sect.~\ref{sec:geometry} is devoted to a geometric interpretation of the WGC. Finally, we offer our conclusions in Sect.~\ref{sec:conclusions}.

\end{section}

\begin{section}{Axion monodromy and Domain Walls}
\label{sec:pheno}

In this section we aim at obtaining constraints on models based on axion monodromy (inflation or relaxation). We begin by pointing out the existence of light domain walls in those models. Interestingly, these are different from the membranes inherent to the KS approach to axion monodromy. They belong purely to the effective field theory regime and do not descend from a higher dimensional gauge theory. We apply the WGC to these low energy domain walls and then discuss the relation of our result to the recent analysis of \cite{Ibanez:2015fcv}. 

\begin{subsection}{Light domain walls}
\label{sub:domainwalls1}

We start by adopting a naive $4d$ effective field theory point of view of axion monodromy models. The Lagrangian of such a model is given by:
\begin{equation}
\mathcal{L}=(\partial\phi)^{2}-V(\phi/f),
\end{equation}
and the inflationary (or relaxion) potential generically consists of a polynomial part and an oscillatory term, e.g.:
\begin{equation}
\label{eq:potential}
V(\phi)=\frac{1}{2}m^{2}\phi^{2}+\alpha\cos\left(\frac{\phi}{f}\right)
\end{equation} 
In writing a cosine term in \eqref{eq:potential}, we are assuming that the axion $\phi$ couples to instantons.\footnote{Even if axions without coupling to gauge theory or stringy instantons exist, the presence of gravitational instantons (see \cite{Montero:2015ofa} and references therein) appears unavoidable.} The results that we will derive in the following subsections rely on the presence of this oscillatory term. Let us remark that, in models of cosmological relaxation, such a contribution is crucial. Furthermore, in models of axion monodromy and relaxation one typically has (and for relaxation actually needs) $\alpha\equiv\alpha(\phi)$, see e.g.~\cite{Baumann:2006th,Flauger:2009ab,Flauger:2014ana}. 

The cosine term generates `wiggles' on top of the quadratic potential. For suitable values of $m, \alpha$ and $f$, namely for $\alpha/(m^2 f^2)>1$, the potential is characterised by the presence of local minima, see Fig.~(\ref{fig:potential1}). In this paper we focus precisely on this case. Slow-roll inflation starts at large values of $\phi$, where the quadratic potential is dominant and there are no local minima. Eventually, the field reaches the region where the wells become relevant and minima appear. We wish to constrain the model with potential \eqref{eq:potential} by focusing on this latter region. `Wiggles' are related to the existence of domain walls: once the inflaton (relaxion) gets stuck in one of the cosine wells, there is a nonvanishing probability to tunnel to the next well, which is characterised by a smaller value of the potential. This happens by the nucleation of a cosmic bubble created by a Coleman-De Luccia instanton, containing the state of lower energy and its rapid expansion.

In order to understand this point, let us adopt a coarse-grained approach: namely, let us consider a model with $V(\phi)$ as in Fig.~(\ref{fig:potential1}) at spatial distances which are larger than the inverse ``mass'' $V^{''}(\phi)^{-1}$. At these distances, $\phi$ is non-dynamical. The ``wiggles'' are therefore invisible, and what remains is a set of points, corresponding to the local minima of the original potential. These points are naturally labelled by an integer index $n$. Therefore the energy of the corresponding configurations is just:
\begin{equation}
\label{eq:energy}
 E\simeq (1/2)m^2\phi_{min}^2\simeq \frac{1}{2}m^2(2\pi f)^2 n^2, \quad n\in\mathbb{Z}.
\end{equation} 
Such a discrete set of vacua can be described in terms of a four-form field strength $F_4=dA_3$.\footnote{Jumping ahead, we note that our use of a four-form flux is therefore different from the approach of \cite{Dvali:2005an, Kaloper:2008fb}, where $F_4$ is introduced as the field strength of a $3$-form which gauges the shift symmetry of the $4d$ dual of $\phi$. This will be discussed in Sec. \ref{sub:ks}.} Indeed, due to gauge symmetry, the theory of a free $3$-form potential in $4d$ has no dynamics (as e.g. in \cite{Brown:1987dd, Brown:1988kg, Bousso:2000xa, D'Amico:2012ji}). 
The points corresponding to the local minima of the original potential for $\phi$ are separated by domain walls. Therefore, the $3$-form lagrangian which provides an effective description of the axion system is:
\begin{equation}
\label{eq:freef4}
\mathcal{L}=\frac{1}{2e^{2}}\int F_{4}^{2}+\int_{DW} A_{3},
\end{equation}
where we have included a phenomenological coupling of $A_3$ to domain walls. It is easy to see that $F_4$ changes by $e^2$ across a membrane, such that $F_4^2/(2e^2)\equiv (1/2)e^2n^2$. By comparison with \eqref{eq:energy}, we find: $e=2\pi m f$.

The tension of the domain wall, i.e. the surface tension of the bubble containing the state of lower energy, can be estimated as the product of the characteristic thickness $b\sim \Delta\phi/\sqrt{V}$ and height of the domain wall $V\sim \alpha$ (see e.g. \cite{Coleman}). One obtains: $T_{DW}\sim \sqrt{V}\Delta\phi\sim \alpha^{1/2} f$. As we make the domain walls lighter, i.e. as we lower the value of $\alpha$, the wiggles become less pronounced, see Fig. (\ref{fig:potential2}).

\begin{figure}[]
  \centering
  \begin{minipage}[b]{0.45\textwidth}
    \includegraphics[width=\textwidth]{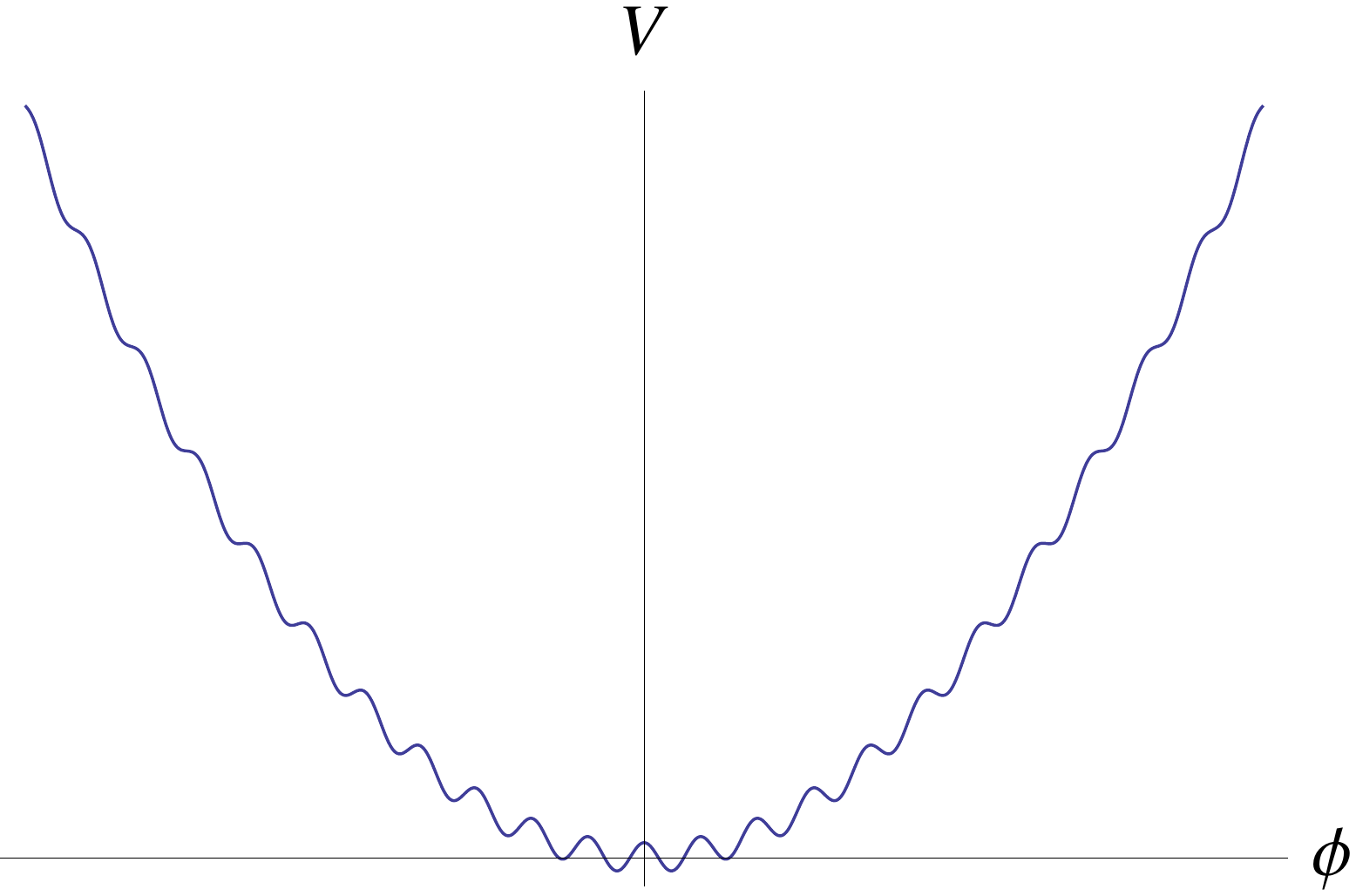}
    \caption{Monodromy potential, as in \eqref{eq:potential}. Here $\alpha/(m^{2}f^{2})\simeq 50$.}
    \label{fig:potential1}
  \end{minipage}
  \hfill
  \begin{minipage}[b]{0.45\textwidth}
    \includegraphics[width=\textwidth]{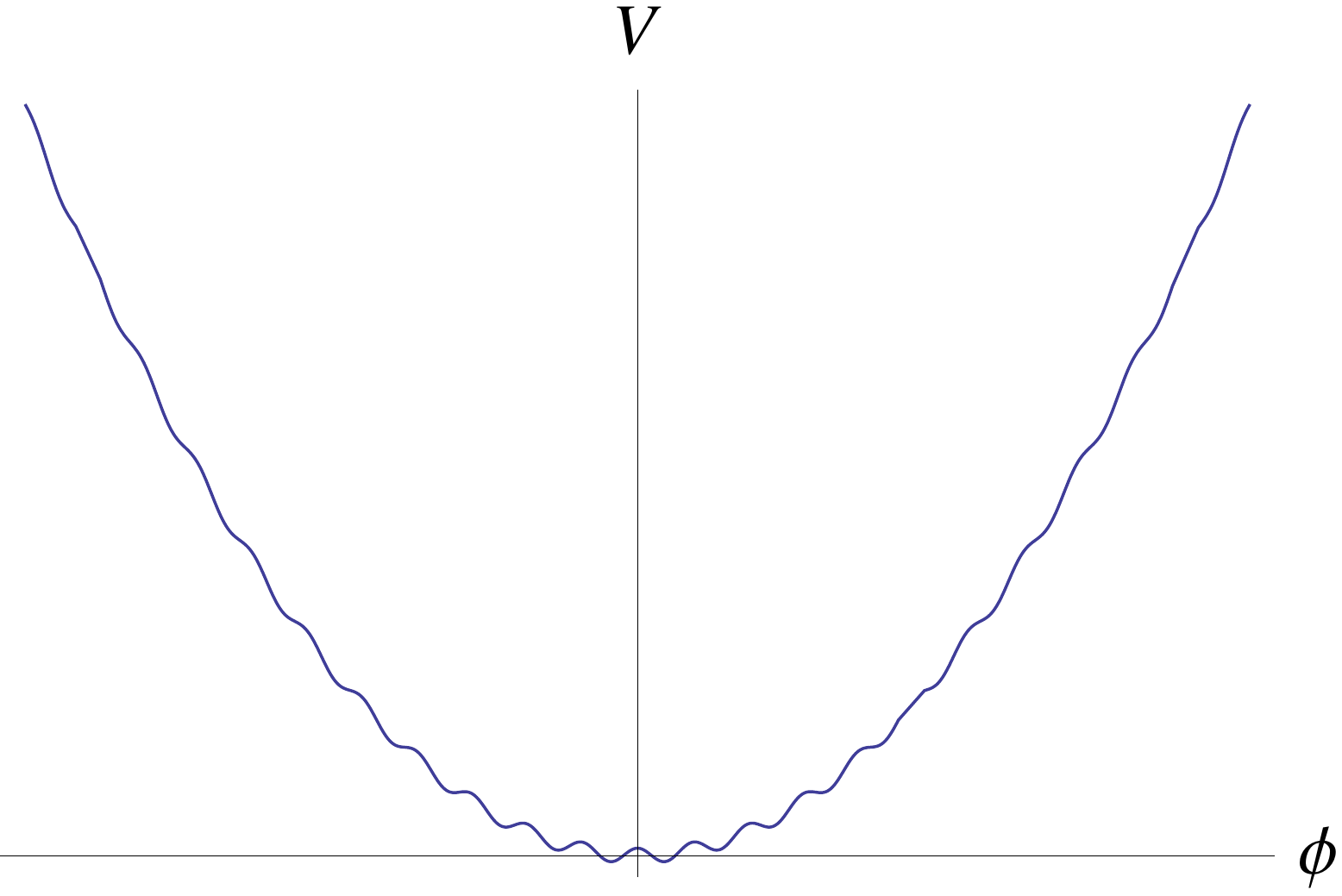}
    \caption{Monodromy potential, as in \eqref{eq:potential}. Here $\alpha/(m^{2}f^{2})\simeq 25$.}
    \label{fig:potential2}
  \end{minipage}
\end{figure}

In order to ensure that the inflaton (or relaxion) can slowly-roll for a sufficiently large distance, one needs to make sure that the height of the wiggles, i.e. the tension of the domain walls, is small enough.\footnote{Nevertheless, the last stages of inflation (or relaxation) may arise as continuous nucleation of cosmic bubbles.}

The crucial point is that lowering the tension of these domain walls goes precisely in the same direction as required by the WGC. Let us recall that, in its original form \cite{ArkaniHamed:2006dz}, the conjecture concerns $4d$ $U(1)$ gauge theories with coupling $e$ and gravity. The electric side of the conjecture requires that a particle of mass $m_{e}$ exists such that: $eM_{pl}/m_{e}\gtrsim 1$. The statement can in principle be extended to any $(p+1)$-form gauge theory in $d$ dimensions, with $p$-dimensional electrically charged objects. The generalisation to domain walls, i.e. $p=2$ in $4d$, is actually not straightforward and may present subtleties (see \cite{Heidenreich:2015nta, Ibanez:2015fcv}). For the moment, we assume that the conjecture is valid for domain walls; we motivate our assumption in detail in Sec. (\ref{sec:geometry}). Therefore, we have the following constraint on the tension and coupling of the domain wall:
\begin{equation}
\text{WGC:}\quad T\lesssim e M_{pl}.
\end{equation}
Applied to inflationary (relaxion) models, this condition leads to $T\lesssim m f M_{pl}$. The conjecture requires a small tension, which is what is needed to have slow-roll inflation (or relaxation). 

Therefore, we are unable to constrain Inflation/Relaxation models by this logic.

\end{subsection}

\begin{subsection}{Constraints from the magnetic WGC}
\label{sec:domainwalls2}

In the previous subsection we have seen that the electric side of the WGC, as applied to light domain walls, does not constrain models based on axion monodromy. However, there exist two versions of the conjecture.\footnote{There exists yet another version of the conjecture, demanding that the states satisfying the WGC are within the validity range of the effective field theory \cite{Heidenreich:2015wga}. In this paper, we do not consider it, since, in string models, this appears not to hold if one identifies the KK scale with the cutoff. Also, there are further variants of the electric version (``strong'',``mild'',``lattice''), which we do not discuss.} The aim of this subsection is to show that the magnetic side imposes a non-trivial constraint on the field range in models of Axion Monodromy (inflation or relaxation).

We start by providing a statement of the magnetic WGC in the form of a constraint on the cutoff $\Lambda$ of a gauge theory. To this aim, let us proceed by dimensional analysis. We consider a $(p+1)$-form gauge theory with coupling $e_{p,d}$ in $d$ dimensions with electrically charged $Dp$-branes and magnetically charged $D(d-(p+4))$ branes. The magnetic WGC simply states that the minimally charged magnetic brane should not be a black brane. The tension of a black brane is $T^{BH}_{d-(p+4)}\sim M_{d}^{d-2}R^{p+1}$, where $R$ is the Schwarzschild radius of the black brane and $M_{d}$ is the Planck scale in $d$ dimensions. The tension of a magnetically charged brane can be estimated by integrating the field strength outside the core, as in the familiar case of the magnetic monopole. In $d$ dimensions and for a $p+1$-form, the coupling has dimensions $[E]^{(p+2)-d/2}$. Therefore: 
\begin{equation}
T_{d-(p+4)}\sim \frac{\Lambda^{p+1}}{e_{p,d}^{2}}.
\end{equation}
The magnetic WGC then requires:
\begin{equation}
\label{eq:magnetic}
\text{magnetic WGC:}\quad T\lesssim T^{BH}\Rightarrow \Lambda^{2(p+1)}\lesssim e^{2}_{p,d}M_{d}^{d-2}.
\end{equation}
Although this derivation does not go through in $4d$ for $p=2$ (since we cannot make sense of $D(-2)$ branes), we conjecture ``by analytical continuation'' in $(p,d)$ that the constraint applies. We therefore obtain:
\begin{equation}
\label{eq:cutoff}
\Lambda\lesssim e^{1/3}M_{Pl}^{1/3}.
\end{equation}
This is the constraint we were after. We will provide more support for it later on.

We now apply \eqref{eq:cutoff} to axion monodromy models. As we have seen in the previous subsection, the coupling $e$ is related to the axion parameters by: $e=2\pi m f$. Therefore, we get the condition $\Lambda\lesssim (2\pi m f M_{pl})^{1/3}$. The relevant constraint is now obtained by requiring that the Hubble scale is below the EFT cutoff, i.e. $H=(V/3M_{pl}^{2})^{1/2}=1/\sqrt{6}\cdot (m/M_{pl}) \phi\lesssim\Lambda$. This gives an upper bound on the field range:
\begin{equation}
\label{eq:constraint}
\frac{\phi}{M_{pl}}\lesssim \left(\frac{M_{pl}}{m}\right)^{2/3}\left(\frac{2\pi f}{M_{pl}}\right)^{1/3}.
\end{equation}
As it stands, the constraint \eqref{eq:constraint}, although non-trivial, represent only a mild bound on the field range. With $m/M_{pl}\sim 10^{-5}$, and $2\pi f/M_{pl}\simeq 1$ one gets $\phi/M_{pl}\lesssim 10^{3}$, which safely allows large field inflation. We expect that our dimensional analysis estimate is modified only by $O(1)$ factors (see Sec. \ref{sec:geometry}). However for models with small $f$ the constraint \eqref{eq:constraint} may become relevant.

It is a generic expectation that, in models of large field inflation, the field range cannot be parametrically large. The discussion of this section confirms this expectation: In the case of axion monodromy, the magnetic side of the WGC limits the field excursions. However, phenomenologically relevant field ranges are allowed. 

Let us now very briefly discuss the corresponding constraint for models of cosmological relaxation \cite{Graham:2015cka} based on monodromy \cite{Ibanez:2015fcv}. In this case, we take our axion $\phi$ to be the relaxion, and couple it to the Higgs field of the standard model. Therefore, the relaxion potential is:
\begin{equation}
V_{\phi}=\frac{1}{2}m^2\phi^2+\alpha_{v}\cos\left(\frac{\phi}{f}\right)+(-M^2+g\phi)|h|^2,
\end{equation}
where $M$ is the cutoff scale and $\alpha_{v}\equiv \alpha(h=v)$. As discussed in \cite{Graham:2015cka}, the following constraints apply to this class of models:
\begin{align}
\label{eq:relconstr}
\Delta\phi &\gtrsim M^2/g \quad \text{to scan the entire range of values of the higgs mass.}\\
H &\lesssim \alpha_{v}^{1/4} \quad \text{to form the low energy barriers.}\\
\label{eq:relconstr1}
H &>M^2/M_{pl} \quad \text{for the energy density to be inflaton dominated.}
\end{align}
Furthermore, the slow roll of $\phi$ ends when the slopes of the perturbative and non-perturbative potential terms are equal, i.e.: $m^2\phi\sim \alpha_{v}/f$. This should happen at a generic point in the range of $\phi$. Hence, from \eqref{eq:relconstr}, $\phi\sim M^2/g$ and thus:
\begin{equation}
\label{eq:rel1}
\frac{m^2M^2}{g}\sim \frac{\alpha_{v}}{f}.
\end{equation}
We can now find the consequences of the magnetic WGC for this class of models. We apply \eqref{eq:cutoff} and require, as in the inflationary case, $H\lesssim \Lambda\lesssim (2\pi m f)^{1/3}M_{pl}^{1/3}$. We express $f$ in terms of $\alpha_v$ by means of \eqref{eq:rel1}. By also imposing \eqref{eq:relconstr1}, we are able to constrain the cutoff $M$ as follows:
\begin{equation}
\label{eq:relbound}
M\lesssim \left(\frac{2\pi g}{m}\right)^{1/8}\alpha_v^{1/8}M_{pl}^{1/2}.
\end{equation}
A similar constraint was given in \cite{Ibanez:2015fcv} (we review this approach in Sec. \ref{sub:ks}), where a more detailed discussion can also be found. Even if $\alpha_v$ is as low as $f_{\pi}^2m_{\pi}^2$ this constraint is not fatal.

Before moving on to the next subsection, we would like to remark on a well-known problem of all the axion models, which also affects our setup. In these models there are always instantons associated to the slowly-rolling axion. If they all contribute to the axion potential, there is no flat direction on which to inflate (relax). It is a non-trivial task to suppress the higher order instantons (our `wiggles'),  and strategies to do so and evade the WGC have been an important focus of recent work (see \cite{Hebecker:2015rya} for a proposal which realises a loophole of the WGC \cite{Rudelius:2015xta, Brown:2015iha}).

Let us now motivate, as promised, our extension of the magnetic WGC to generic $p$-dimensional objects.

\end{subsection}

\begin{subsection}{String Theory and the WGC}
\label{sub:motivation}

It has been suggested in \cite{Ibanez:2015fcv} that there is no magnetic side of the conjecture for domain walls, a statement which conflicts with our previous discussion. Here we would therefore like to motivate our use of the magnetic WGC. From now on, we work in units where $M_{pl}\equiv 1$.

From the point of view of string theory, there are two possible ways of satisfying the electric WGC. On the one hand, string compactifications may provide light objects whose tension and coupling satisfy the inequality $T\lesssim e$. However, $Dp$-branes in $10D$ are extremal, i.e. they marginally fulfil the WGC. Under compactifications, the resulting objects are not guaranteed to be extremal, unless SUSY is preserved. Therefore, it is not clear whether objects arising from string compactifications could violate the WGC.

On the other hand, there exists another mechanism by which the conjecture can be satisfied in string compactifications: It is the presence of a maximal scale up to which a 4d effective field theory description is valid. In many cases such a cutoff is set by the KK scale $M_{KK}\sim 1/R$, where $R$ is the typical length scale of the compactification manifold. Above $M_{KK}$, one has to work with the full 10D theory. In particular, if the tension of the objects descending from string theory is larger than $M_{KK}$, then these objects simply do not exist in the low energy effective field theory. Therefore, by lowering the KK scale, one can ensure that the WGC is not violated, by simply removing the dangerous objects from the spectrum of the low energy theory. A low cutoff is precisely what is required by the magnetic side of the WGC for a weakly coupled theory. 

Explicitly, consider a $q$-dimensional object descending upon compactification from a $p$-dimensional brane in 10D. The ratio between its tension and the appropriate power of the KK scale is given by: $\tau_{q}/M_{KK}^{q+1}\sim M_{s}^{p+1}R^{p+1}/g_{s}$. We are assuming that we are in a controlled regime, i.e. either $g_s<1$ or $R>1$ or both. Therefore as $R$ increases the corresponding object simply disappears from the 4d theory. 

The bottomline of this discussion is that, in many cases, string theory satisfies the WGC by imposing a low cutoff to the 4d effective field theory, not by providing objects which are light enough. In other words, setups of string compactifications satisfy the magnetic side of the WGC and, as a consequence, the electric side as well.

This is the reason why we think that the magnetic constraint is the more fundamental conjecture among the two version of the WGC. Therefore, we assume that the magnetic WGC is valid for any p-form, and in particular for domain walls. 

Recently, the electric WGC has been applied to another class of membranes in the context of realisations of axion monodromy models \`{a} la Kaloper-Sorbo (KS) (see \cite{Kaloper:2008fb} for the KS proposal, \cite{Ibanez:2015fcv} for the recent developments) . In the next subsection we describe the relation of this work to our findings. 

\end{subsection}

\begin{subsection}{Relation to domain walls \` {a} la Kaloper-Sorbo}
\label{sub:ks}

We begin by reviewing the strategy of \cite{Brown:2015iha, Ibanez:2015fcv} to constrain nucleation rates in models based on axion monodromy. In this subsection, we follow the notation of \cite{Ibanez:2015fcv}, which differs from the one used in the previous subsections.

The KS proposal \cite{Kaloper:2008fb} to implement monodromy models in a $4d$ setup is to introduce a 3-form gauge potential $A_{3}$ and to couple the corresponding 4-form field strength to the axion:
\begin{equation}
\label{eq:kaloper}
\mathcal{L}= -\frac{1}{2}\left(\partial_{\mu}\phi\right)^{2}-\frac{1}{2}|F_{4}|^{2}+g\phi F_{4},
\end{equation}
where $|F_{p}|^{2}\equiv\frac{1}{p!}F_{\mu_{1}\dots\mu_{p}}F^{\mu_{1}\dots\mu_{p}}$. Notice that this setup is different from the dual picture that we have described in Sec. (\ref{sub:domainwalls1}). We used just one scalar field theory with a discretuum of vacua, which corresponds to a gauge theory with the same discretuum of vacua. By contrast, the lagrangian in \eqref{eq:kaloper} consists of a scalar field theory (first term) and a gauge theory (second term), each with its own set of vacua. The third term couples these two theories. The potential $A_{3}$ couples to fundamental $4d$ domain walls via $S\sim q\int_{2-branes} A_{3}$. The field strength $F_{4}$ varies across the membranes and is quantised in units of the membrane charge, i.e. $F_{4}=nq~(\star 1)$. A shift in the value of $F_{4}$ is a part of the residual gauge symmetry of the KS lagrangian. Under this symmetry, also the scalar field shifts:
\begin{equation}
\label{eq:gauge}
\phi\rightarrow \phi+2\pi f, \quad nq\rightarrow (n-k)q, \quad n,k\in\mathbb{Z},
\end{equation}
with the consistency condition $2\pi fg=kq$, and $f$ being the axion periodicity. Due to this residual gauge symmetry, we are left with only one set of vacua, labeled by one integer.

The quadratic potential for $\phi$ arises from integrating out the field strength $F_{4}$:
\begin{equation}
V_{KS}=\frac{1}{2}(nq+g\phi)^{2}.
\end{equation}
The crucial point is that each value of $n$ corresponds to a different branch of the potential. The gauge symmetry \eqref{eq:gauge} provides a way to identify these branches. In this sense, crossing a membrane corresponds to an alternative way to move one step down in the potential. This is different from rolling over or tunneling through a ``wiggle'' of Fig. (\ref{fig:potential1}). The KS membranes can potentially spoil the slow-roll behavior allowed by small ``instanton-induced wiggles''. As usual, the probability for such tunneling events is described in terms of a nucleation rate for the corresponding bubbles.

Since this probability is exponentially suppressed, one might wonder whether this effect represents a concern for Axion Inflation. The nucleation rate $\Gamma$ is given by $e^{-B}$, where $B\sim T/H^{3}$ in the relaxion regime (see \cite{Coleman:1980aw}).

In \cite{Brown:2015iha}, the authors show that a strong suppression of the nucleation rate requires a violation of the WGC.
More recently, in \cite{Ibanez:2015fcv}, the authors follow the same direction to constrain models of relaxion monodromy. In this case, the WGC requires $T\lesssim 2\pi fg$. By requiring $B>1$, the authors obtain a constraint which is similar to \eqref{eq:relbound}. In particular, the parametric dependence on $\alpha_v$ is the same. Furthermore, the authors of \cite{Ibanez:2015fcv} obtain a stronger constraint by requiring that $B>N$, where $N$ is the number of e-folds. This requirement arises from demanding that there are no domain walls in the part of the universe created during the above $N$ e-folds. Such a constraint cannot be obtained by using our low energy wiggles, because the latter arise only much later, when inflation is in its last stages.

Applied to inflationary models, $T\lesssim e M_{pl}$ and $T\gg H^{3}$ lead to the same constraint that we have found in \eqref{eq:constraint}. However, we have obtained it by using a different, arguably simpler, effective field theory point of view, based on the magnetic, rather than the electric WGC. Notice also that the objects that we have described in Sec. (\ref{sub:domainwalls1}) can be naturally lighter than the KS membranes. 

We have seen that the Kaloper-Sorbo procedure consists in gauging an axionic theory by a $3$-form potential. The original theory of a free $3$-form potential has domain walls to which the WGC can be applied. However, gauging corresponds to a discontinuous, qualitative change of the model.\footnote{In particular, the conceivable limiting procedure of taking the gauge coupling to zero and hence going from the gauged to the ungauged case is forbidden by the WGC itself. Also, in string constructions gauging often corresponds to (necessarily discrete) changes in the flux configuration or even in the geometry of the compact space.}

It is hence not clear whether the relevant parameters, i.e. the tension and the coupling, and therefore the consequences of the WGC, remain unchanged. In particular, it may be hard to check the changes of the coupling and tension of the domain walls, since the dualisation procedure described in \cite{Dvali:2005an} does not always lead to an explicit determination of the $F_4$ lagrangian. Therefore, it is desirable to work with constraints which do not appeal to the situation  \emph{before} gauging. Crucially, \emph{after} the gauging both the fundamental KS domain walls and the `wiggle-induced' effective domain walls are present.

We are then left with two possibilities: The first is that the electric WGC has to be separately satisfied by both the KS and by the effective domain walls described in this paper. In this case the constraint given in \cite{Ibanez:2015fcv} and based on the electric WGC for the (heavier) KS domain walls applies. The same constraint arises as a consequence of the magnetic WGC. Everything is consistent and the present paper provides an alternative derivation of the same constraint.

The other possibility is that the electric WGC needs to be satisfied only by the lightest domain walls. These are the effective domain walls, but due to their lightness no interesting constraint arises. The heavier KS domain walls provide no further constraints. Thus, the magnetic WGC provides the only useful constraint, as described in this paper.

Our conclusion is that in both cases the field range is constrained according to \eqref{eq:constraint}. In the first case the latter comes from the electric side applied to KS membranes, as explained in \cite{Brown:2015iha,Ibanez:2015fcv}, and from the magnetic side applied to ``wiggles'' membranes. In the second point of view, which we adopt in this paper, no UV information on the origin of the gauge theory is required and \eqref{eq:constraint} follows only from the magnetic WGC. 

\end{subsection}

\end{section}

\begin{section}{The WGC as a geometric constraint}
\label{sec:geometry}

In this section we want to address the extension of the WGC to domain walls. We will do so in the framework of 10D string theory compactified on a CY manifold. 

\begin{subsection}{Previous approaches and our perspective}

In \cite{Heidenreich:2015nta}, the authors provide the following statement of the WGC for any $p$-form in $d$ dimensions:

\begin{equation}
\label{eq:wgc}
\left[\frac{\alpha^2}{2}+\frac{p(d-p-2)}{d-2}\right]T_{p}^{2}\leq e^{2}_{p:d}q^{2}M_{d}^{d-2}.
\end{equation}
In the absence of a dilaton background, the inequality is degenerate for $p=0$ (axions) and $p=d-2$ (strings). Moreover, for $p=d-1$, i.e. for domain walls, the inequality cannot be satisfied, as already pointed out in \cite{Heidenreich:2015nta}. Therefore, one may worry that there is no statement of the electric WGC for domain walls.

An idea to extend the WGC to generic $p$-dimensional objects, as noticed in \cite{Ibanez:2015fcv}, is to use string dualities. This follows very closely the strategy of \cite{Brown:2015iha}, where the conjecture is extended to axions and instantons. In that case, the authors consider type IIB on a CY 3-fold with $D1$ branes and their associated $C_{2}$ gauge potential. Wrapping the branes on $2$-cycles and compactifying to 4d, one obtains a theory of $C_{2}$ axions and $D1$ instantons. This type IIB theory is then T-dualised to type IIA with $D2$ branes and their associated $C_{3}$ potential. Since this theory is strongly coupled, one actually uses the $M$-theory picture, introducing a further compact dimension. Again, by wrapping the branes around $2$-cycles and compactifying, one obtains a $5d$ theory with a $U(1)$ gauge field and $M2$ particles. This is the original content of the WGC, which can therefore be applied to this particular 5D setting. Finally, one can translate the constraints obtained on the particles/vector fields side to the axion/instanton side, by using the T-duality relations between IIA and IIB couplings and mass scales. 

In \cite{Ibanez:2015fcv}, the authors propose to implement the very same idea to constrain domain walls. Starting with a 10D theory with $p=d-1$ objects, they propose to T-dualise twice along directions transverse to the branes, so that the dual theory is of the same type but with $p=d-3$ branes. One can then apply \eqref{eq:wgc} to the latter setup, then translate the constraints to the domain walls side.

We agree with the authors of \cite{Ibanez:2015fcv} that the apparent problems of the WGC for domain walls disappear when considering them in a string theory setup. Notice that the dualisation procedure works for any $p$-dimensional object in $10$ dimensions reduced to a $q$ dimensional object in $d$ dimensions. Indeed the moduli of the theory, i.e. the compactification radius and the string coupling, disappear from the charge-to-tension ratio on both sides of the duality. Were this not the case, we would not be able to extract a sensible constraint on the objects in the $4d$ theory. 

This property suggests that the WGC in 10D string theory can be phrased as a constraint on some geometrical data of the particular compactification manifold, independently of the specific $p$-dimensional object. Once the geometry of the compactification manifold is constrained, one can extract the consequences for any other $q$-dimensional object in the theory. This is the novel point of this section. Our focus in this section is the electric statement.

Our approach implies that there is no need of T-dualising in order to extend the WGC to objects other than $4d$ particles. In the next subsection, we will verify this statement focusing on the case of domain walls. Let us therefore outline the strategy to extend the conjecture to any $p$-dimensional object, without using dualities. One starts with a type IIB setting with $Dp$ branes wrapped on $p$-cycles of the internal manifold $X$. Upon compactification, this leads to a 4d theory of particles and gauge fields. One then applies the standard WGC to this setting: the result is a constraint on the metric on the space of $p$-cycles in $X$. For example, in \cite{Brown:2015iha} the authors obtain a constraint on $K_{ab}\sim \int w_{a}\wedge\star w_{b}$, where $w_{a}$ is a basis of $H^{2}(X,\mathbb{Z})$. Once this constraint is obtained, it is valid for any brane setup on the same CY. One can then consider $Dq$ branes, with $p\neq q$ wrapped on the same $p$ cycles and obtain inequalities for the tension and couplings of the $4d$ theory derived by compactification on $X$.

\end{subsection}

\begin{subsection}{Computation}

Following our discussion, we now perform an explicit computation to prove our claim. We first focus on obtaining particles in $d=4$. As a starting configuration we choose type IIB with D3 branes compactified on a CY 3-fold $X$. Other choices are equally valid. We work in the conventions of \cite{Polchinski}. The relevant 10d action reads:

\begin{equation}
\label{eq:IIBD3}
S_{10}\supset \frac{1}{2\kappa_{10}^{2}}\int_{M^{10}}\left[\frac{1}{g_{s}^{2}}R_{10}\star\mathbf{1} -\frac{1}{4}F_{5}\wedge\star F_{5}\right]+\mu_{3}\int_{D3} C_{4}
\end{equation}
where $\kappa_{10}^{2}=(1/2)(2\pi)^{7}\alpha'^4$ and $\mu_{3}=2\pi(4\pi^{2}\alpha')^{-2}$. Now let us perform dimensional reduction, by wrapping the $D3$ on 3-cycles of $X$. We focus on the gauge kinetic term. We consider a symplectic basis $w_{i}=(\alpha_{a},\beta^{b})$ of $H^{3}(X,\mathbb{Z})$, i.e. s.t.:
\begin{equation}
\label{eq:symplectic}
\int_{X}\alpha_{a}\wedge\beta^{b}=\delta_{a}^{b},
\end{equation}
and the other intersection numbers vanish. By Poincaré duality, one can define the integral charges:
\begin{equation}
q_{i}^{k}=\int_{\Sigma_{k}}w^{i}=\int_{X}w^{i}\wedge w^{k},
\end{equation}
where $\Sigma_{k}$ is a $3$-cycle in $X$ and $w^{k}$ is its dual form. By \eqref{eq:symplectic}, these charges are either vanishing or unit.

The 4d action is obtained by expanding the five-form flux and the four-form potential in terms of the symplectic basis of $H^{3}(X,\mathbb{Z})$:
\begin{equation}
\label{eq:decomposition}
F_{5}=\sum_{i=1}^{N}F_{2}^{i}(x)\wedge w_{i}(y), \quad C_{4}=\sum_{i=1}^{N}A^{i}_{1}(x)\wedge w_{i}(y),
\end{equation}
then integrating over $X$. Here $N$ is the number of $3$ cycles of $X$. $D3$-branes wrapping a $3$-cycle $\Sigma$ generate particles in the $4d$ theory. For the moment being, let us focus on one such cycle. We will later extend our results to particles descending from different cycles.
Let us introduce the metric on the space of $3$-forms:
\begin{equation}
\label{eq:metric}
K_{ij}\equiv \int_{X}w_{i}\wedge\star w_{j}.
\end{equation}
Before moving to the $4d$ theory, an important remark is in order. This concerns the self-duality of $F_{5}$, i.e. $\star{F}_{5}=F_{5}$. This constraint cannot be implemented at the level of the $10d$ action \eqref{eq:IIBD3}. Therefore one actually starts with a more general theory where $F_{5}\neq \star F_{5}$. Nevertheless, the kinetic term in \eqref{eq:IIBD3} is normalised with a prefactor $1/4$ instead of $1/2$, as will be appropriate after self-duality is imposed. To obtain consistent $10d$ equations of motion, the coupling in \eqref{eq:IIBD3} should actually read:
\begin{equation}
\label{eq:fullcoupling}
S_{10}\supset \frac{\mu_{3}}{2}\int_{D3} C_{4} +\frac{\mu_{3}}{2}\int_{\tilde{D3}}\tilde{C}_{4},
\end{equation} 
where at the moment different branes source the dual potentials (see also \cite{Giddings:2001yu}, footnote n. 6). Self-duality can then be consistently imposed at the level of the $10d$ equations of motion, derived from this action. This goes together with identifying $D3$ and $\tilde{D3}$.

We now consider the $4d$ theory descending from dimensional reduction. In $4d$ there are certain constraints on the field strengths $F_{2}^{i}$ and $\tilde{F}_{2}^{i}=\star F_{2}^{i}$ descending from self-duality of $F_{5}$ in $10d$. For the sake of our analysis, we first focus on the set of unconstrained $F_{2}^{i}$, exactly as we we did with $F_{5}$ in $10d$. The $4d$ action then reads:
\begin{equation}
\label{eq:4daction}
S_{4}\supset \frac{M_{pl}^{2}}{2\cdot 4}\int_{M^{4}}\frac{g_{s}^{2}}{V_{X}}K_{ij}F_{2}^{i}\wedge\star F_{2}^{j}+q^{\Sigma}_{i}\frac{\mu_{3}}{2}\int_{0-brane} A_{1}^{i}.
\end{equation}
Here $M_{pl}^{2}=V_{X}/\kappa_{10}^{2}g_{s}^{2}$ is the $4d$ Planck mass. 
The equations of motion arising from \eqref{eq:4daction} read
\begin{equation}
d\star F^{i}K_{ij}=\frac{2V_{X}}{M_{pl}^{2}g_{s}^{2}}\mu_{3}q^{\Sigma}_{i}d j_{\text{0-brane}}.
\end{equation}
From the latter, it is clear that only a certain linear combination of gauge fields is sourced by the particle with charge $q^{\Sigma}_{i}$. To make this visible in the 4d action, we define the field $A_{1}$ and its field strength $F_{2}=dA_{1}$ by
\begin{equation}
A_{1}^{i}\equiv A_{1}K^{ij}q_{j}^{\Sigma}.
\end{equation}
In terms of $A_{1}$ and $F_{2}$ the 4d action reads
\begin{equation}
\label{eq:s41}
S_{4}\supset \frac{M_{pl}^{2}}{2\cdot 4}|\mathbf{q}^{\Sigma}|^{2}\frac{g_{s}^{2}}{V_{X}}\int_{M^{4}}F_{2}\wedge\star F_{2}+|\mathbf{q}^{\Sigma}|^{2}\frac{\mu_{3}}{2}\int_{0-brane} A_{1},
\end{equation}
where $|q^{\Sigma}|^{2}\equiv K^{ij}q^{\Sigma}_{i}q^{\Sigma}_{j}$. Now we consider the realistic setup where the dual field strength $\tilde{F}_{2}$ and its associated $\tilde{A}_{1}$ are also included. Therefore we add to \eqref{eq:4daction} the action:
\begin{equation}
\label{eq:s42}
\tilde{S}_{4}\supset \frac{M_{pl}^{2}}{2\cdot 4}\int_{M^{4}}\frac{g_{s}^{2}}{V_{X}}K_{ij}\tilde{F}_{2}^{i}\wedge\star \tilde{F}_{2}^{j}+\tilde{q}^{\Sigma}_{i}\frac{\mu_{3}}{2}\int_{\widetilde{0-brane}} \tilde{A}_{1}^{i}
\end{equation}
where $\tilde{A}_{1}$ and $\tilde{q}_{i}^{\Sigma}$ are analogous to $A_{1}^{i}$ and $q^{i}_{\Sigma}$. The coupling term can be obtained from dimensionally reducing \eqref{eq:fullcoupling}. Finally we relate $F_{2}^{i}$ and $\tilde{F}^{i}_{2}$ by dimensionally reducing $10d$ self-duality of $F_{5}$. In particular, $10d$ self-duality implies $F_{2}^{J}=\star_{4}F_{2}^{K}H_{K}^{J}$, where the matrix $H_{K}^{J}$ is defined by $\star_{6}w_{K}=H_{K}^{J}w_{J}$. In imposing this constraint, we also identify the branes sourcing $A_{1}^{i}$ and $\tilde{A}_{1}^{i}$. Thus, adding \eqref{eq:s41} and \eqref{eq:s42} corresponds to a doubling of the action \eqref{eq:s41}. Therefore, the final theory which we will constrain via the WGC has action
\begin{equation}
S_{4}\supset \frac{M_{pl}^{2}}{2\cdot 2}|\mathbf{q}^{\Sigma}|^{2}\frac{g_{s}^{2}}{V_{X}}\int_{M^{4}}F_{2}\wedge\star F_{2}+|\mathbf{q}^{\Sigma}|^{2}\mu_{3}\int_{0-brane} A_{1}.
\end{equation}
In order to extract the $4d$ gauge coupling, we normalise the gauge potential. Finally, we obtain:
\begin{equation}
\label{eq:4def}
S_{4}\supset \frac{1}{2e^{2}}\int_{M^{4}}F_{2}\wedge\star F_{2} +\int_{0-brane} A_{1},
\end{equation}
where we have kept the same notation for the normalised fields and the $4d$ gauge coupling is defined as:
\begin{equation}
\label{eq:coupling}
e^{2}=\frac{2V_{X}\mu_{3}^{2}|q^{\Sigma}|^{2}}{M_{pl}^2 g_{s}^{2}}.
\end{equation}
The result of this procedure is therefore a $4d$ theory of a $U(1)$ gauge field with coupling \eqref{eq:coupling}. The particle descending from the $D3$ brane wrapped on $\Sigma$ has mass $M^{\Sigma}=(T_{3}/g_{s})\int_{\Sigma}\star \mathbf{1}=(T_{3}/g_{s})V^{\Sigma}$, and $T_{3}=\mu_{3}$.

We are now ready to apply the WGC to the $4d$ theory defined by \eqref{eq:4def} with particles of mass $M^{\Sigma}$:
\begin{equation}
\label{eq:wgcparticles}
\frac{eM_{pl}}{M^{\Sigma}}\geq \frac{\sqrt{2}}{2}\Rightarrow \frac{V_{X}^{1/2}|q^{\Sigma}|}{V^{\Sigma}}\geq \frac{1}{2}.
\end{equation}
Before moving to the case of domain walls, let us pause to extract the full meaning of \eqref{eq:wgcparticles}. The WGC for particles arising from a string compactification translates into a purely geometric constraint on the size and intersections of the cycles of the manifold, in this case $3$-cycles. Crucially, all couplings and $4d$ scales have disappeared from the final statement. Despite the presence of volume factors, the charge-to-mass ratio is independent on any rescaling of the $6d$ metric $\tilde{g}_{mn}$. This statement is actually true for any $p$-cycle: indeed the metric $K_{ij}$ on the dual space of $p$-forms contains $(3-p)$ powers of the $6d$ metric, so the numerator scales as $\tilde{g}_{mn}^{p/2}$, but so does the denominator. 

The conclusion is as follows: the procedure that we have followed works for any $p$-dimensional object and associated field strength defined on a chosen manifold $X$ and dimensionally reduced to a $q$ dimensional object in $4d$. In particular, \eqref{eq:wgcparticles} is a constraint on the $3$-cycles of $X$. As such, it can be applied applied to any other 4d object descending from any $p$-brane on $X$ wrapped on the same $3$-cycles. 

We are particularly interested in constraining $4d$ domain walls. In order to apply our previous result, we study the case in which the membranes arise from compactifications of type IIB string theory with $D5$ branes wrapped on $3$-cycles. The action is obtained by simply replacing the $D3$ branes with $D5$ branes in \eqref{eq:IIBD3}:
\begin{equation}
S_{10}\supset \frac{1}{2\kappa_{10}^{2}}\int_{M^{10}}\left[-\frac{1}{2}F_{7}\wedge\star F_{7}\right]+\mu_{5}\int_{D5} C_{6}+S_{DBI},
\end{equation}
with $\mu_{5}=\mu_{3}/(2\pi\alpha')$. Dimensional reduction to $4d$ goes as in the previous case, therefore we do not repeat the computation. The $4d$ action reads:
\begin{equation}
S_{4}\supset \frac{1}{2e_{DW}^{2}}\int_{M^{4}}F_{4}\wedge\star F_{4} +\int_{D2} A_{3}
\end{equation}
with:
\begin{equation}
e_{DW}^{2}=\frac{2V_{X}\mu_{5}^{2}|q^{\Sigma}|^{2}}{M_{pl}^2 g_{s}^{2}}.
\end{equation}
The tension of the $4d$ domain wall is: $T_{DW}=T_{5}/g_{s} V^{\Sigma}$. The charge-to-tension ratio is:
\begin{equation}
\label{eq:ratiodw}
\frac{e_{DW}M_{pl}}{T_{DW}}=\frac{(2V_{X})^{1/2}|q^{\Sigma}|}{V^{\Sigma}}.
\end{equation}
As expected, \eqref{eq:ratiodw} is the same as \eqref{eq:wgcparticles}. Therefore the WGC constraint on particles translates into the following inequality for the charge-to-tension ratio of domain walls:
\begin{equation}
\label{eq:wgcdomainwalls}
\text{WGC:}\quad \frac{e_{DW}M_{pl}}{T_{DW}}\geq \frac{\sqrt{2}}{2}.
\end{equation}
This is the result we were after, namely a WGC for domain walls.  

One can give a general inequality for a $(q+1)$-dimensional object in $d$ dimensions descending from a $s$-brane wrapped on a $(s-q)$ cycle of a CY $X$, by relating its charge-to-mass ratio to that of particle descending from a $p$-brane wrapped on the same $(s-q)$ cycle. For consistency $s-q=p$. The WGC then states that the charge-to-tension ratio of the $D(q)$-brane must satisfy the condition:
\begin{equation}
\label{eq:generalratio}
\frac{e_{p}M_{pl}}{T_{p}}\geq \sqrt{\frac{d-3}{d-2}}.
\end{equation}
Finally, let us generalise our results to the case of $N$ cycles $\Sigma_{k}, k=1,\dots, N$. Correspondingly, we have a set of charge vectors $\mathbf{q}^{\Sigma_{k}}$. These vectors belong to $\mathbb{R}^{N}$ equipped with metric $K_{ij}$ defined as in \eqref{eq:metric}. With the same notation as above, consider $Dp$-branes wrapped on $p$-cycles of a CY manifold. These lead to particles in d dimensions with mass $M_{k}$. The Convex Hull Condition (CHC) for the $p$-cycles reads:

\bigskip
\noindent \emph{The convex hull spanned by the vectors $\mathbf{z}^{k}\equiv \frac{V_{X}^{1/2}\mathbf{q}^{\Sigma_{k}}}{V^{\Sigma_{k}}}$, must contain the ball of radius $r=\sqrt{\frac{d-3}{d-2}}$.}
\bigskip

\noindent Now consider a $q$-brane in $d$ dimensions obtained by wrapping a $D(s)$-brane on $p$-cycles of the same CY. The tension and the charge vectors of the $(q+1)$-dimensional objects are respectively $T_{q}^{k}$ and $e_{q}\mathbf{q}^{\Sigma_{k}}$, where $e_{q}$ is the prefactor in $(1/2)(1/e_{q}^{2})\int K_{ij}F^{i}_{q+2}\wedge\star F^{j}_{q+2}+q_{i}^{\Sigma}\int_{\Sigma_{k}} A_{q+1}^{i}$ in the effective theory. Assuming the CHC for particles, we obtain the following statement for the $q$-branes:

\bigskip
\noindent \emph{The convex hull spanned by the vectors $\mathbf{Z}^{k}\equiv\frac{e_{q}\mathbf{q}^{\Sigma_{k}}M_{d}}{T^{k}_{q}}$ must contain the ball of radius ~$r_{q}=\sqrt{\frac{d-3}{d-2}}$.}
\bigskip

It is important to remark that \eqref{eq:wgcdomainwalls} has been obtained without using any string duality: the WGC for particles imposes a constraint on the geometry of CY three cycles. This constraint, applied to objects derived from any $p$-brane in the 10d setup, translates to a corresponding WGC for these particular objects. 
This line of reasoning can be applied also to the case of axions and instantons. In that case one starts from a $Dp$ brane wrapped on $p$ cycles, then considers $D(p-1)$ branes wrapped on the same cycles. Obviously this requires a change in the theory, e.g. from type IIB to type IIA/M-theory on the same CY. However, the constraints obtained in the IIB setting are still just geometric constraints on $p$-cycles of the CY, therefore there is no need of performing a duality between the two theories. It is sufficient to consider a type IIA/M-theory setup with the appropriate branes, and impose on this setup the previously determined geometric constraint. It would be interesting to think about manifolds with backreaction and fluxes. In this case, the transition from IIA and IIB (or other setups) would not be so straightforward.

\end{subsection}
\end{section}

\begin{section}{Conclusions}
\label{sec:conclusions}

In this paper we have investigated two different aspects of the Weak Gravity Conjecture. Firstly, we have discussed its consequences for models based on Axion Monodromy (Inflation and Relaxation). Secondly, we have provided a geometric interpretation of the conjecture in the framework of string compactification. We now provide a detailed summary of our results.

In the first part of this paper, we have adopted an effective field theory point of view. Namely, given a certain scalar potential, we have tried to constrain its use in models of monodromy inflation. In particular,  inflaton (relaxion) potentials in models of Axion Monodromy are characterised by the presence of `wiggles' on top of a polynomial potential. The resulting local minima imply the existence of $4d$ domain walls. This is more evident by using an effective description in terms of a four-form flux, whose value changes across these membranes.

We assumed that the WGC can be extended to domain walls. In our setup, its electric version gives an upper bound on the tension of the $4d$ membranes. Crucially, this condition agrees with what is required to realise slow-roll: as the tension decreases, the height of the `wiggles' decreases and slow roll can be seen as a continuous nucleation of cosmic bubbles. Therefore, we conclude that, in this logic, the electric WGC does not constrain models of axion monodromy (Inflation and Relaxation). 

For this reason, we focused on the constraints imposed by the magnetic side of the WGC, which we stated as an upper bound on the cutoff of a generic $(p+1)$-form gauge theory  (in the spirit of \cite{ArkaniHamed:2006dz}). We then applied the condition to inflationary models, i.e. we required $H\ll \Lambda$. This gives a non-trivial constraint on the field range: $\phi\lesssim m^{-2/3}f^{1/3}M_{pl}^{4/3}$. The latter however allows for large field displacements, but forbids models with a small decay constant.

We then discussed our extension of the magnetic WGC. We argued that string theory lowers the KK scale to fulfil the WGC for objects which descend from compactifications of string theory with $Dp$-branes, rather than making them light enough. As a consequence, heavy ``stringy'' objects, which could potentially violate the WGC are confined above the cutoff $M_{KK}$. Therefore they do not exist from an effective field theory point of view. Of course, low energy light objects are allowed, as is the case for our domain walls.  Consequently, the electric side is automatically satisfied. We suggest that the magnetic WGC should be seen as the fundamental constraint among the different versions of the WGC.

Recently, the electric WGC has been applied to membranes arising from the realisation of Axion Monodromy \' {a} la Kaloper-Sorbo (KS), in the context of new realisations of relaxion models \cite{Ibanez:2015fcv}. When the tension of these membranes decreases, the probability of tunneling to another branch of the potential increases. Such a transition can spoil slow-roll, as it corresponds to discrete ``jumps'' in the axion trajectory. The requirement that the tunneling rate is suppressed parametrically leads to the same constraint on the field range that we obtained by studying the domain walls arising from `wiggles' in the axion potential.

However, KS membranes are different from the low energy domain walls described in this paper. This may have implications for the various constraints.

There are, in fact, two possibilities. On the one hand, one could impose the WGC separately on the two classes of membranes. In this case, the  constraints given in \cite{Ibanez:2015fcv} for relaxion models apply and can be extended to inflationary models. The magnetic WGC applied to the low energy domain walls gives the same constraint.

On the other hand, it is possible that only the lightest domain walls have to satisfy the WGC. In this case, the electric WGC applied to the low energy domain walls does not give any constraint. By contrast, the magnetic side gives a bound on the field range, hence playing a central role. As discussed in this paper, there are reasons related to the KS gauging which make this second possibility relevant.

In the second part of this paper, we worked in the framework of string compactifications. We started with 10D type IIB with $D3$ branes and compactified to $4d$ by wrapping the branes around $3$-cycles of a CY manifold. Therefore, we obtained particles and gauge fields in $4d$. We applied the original WGC to this setup. Very interestingly, the final constraint does not depend on the couplings and moduli of the 10D setup. The electric WGC translates into a purely geometric constraint on the size and intersection of the $3$-cycles of the CY. The same happens for any $p$-dimensional object wrapped on the same $3$-cycles. Therefore, by constraining the geometry of those cycles through the $D3$/particles case, we obtain a WGC for any $p-3$-dimensional object in $4d$ arising from compactification of type IIB with $Dp$-branes wrapped on $3$-cycles. In particular, by taking $p=5$ we obtain the WGC for $4d$ domain walls. Crucially, we do so without the use of string dualities.

The same procedure applies to any $p$-dimensional object wrapped on some $q$-cycle of a CY, to obtain a $p-q$-dimensional object in $4d$. Therefore, our approach provides a simple strategy to extend the electric WGC to any $q$-dimensional object, without the use of string dualities. 

Let us close our discussion with observing two further consequences implied by the constraint on the tension of the low-energy domain walls from the WGC. We note firstly, that we get a fundamental upper bound on the size of resonant oscillating non-Gaussianity induced by the `wiggles' in the scalar potential. Following the analyses of~\cite{Chen:2008wn,Flauger:2009ab}, the magnitude $f_{NL}^{res.}$ of this type of non-Gaussianity with an oscillating shape in $k$-space is approximately given by $f_{NL}^{res.}\sim b M_{pl}^{3}/(f\phi)^{3/2}$. Here, $b = \alpha / (m^2f\phi)$ denotes the `monotonicity' parameter of the scalar potential with `wiggles' ($b<1$ corresponds to $V'>0$ for $\phi>0$). We can rewrite this as $b=\alpha f^2/(m^2f^3\phi)\sim T_{DW}^2/(m^2f^3\phi)< m^2f^2M_{pl}^2/(m^2f^3\phi)=M_{pl}^2/(f\phi)$ where the inequality arises from the WGC $T_{DW}< eM_{pl}=mfM_{pl}$. Hence, we get a bound $f_{NL}^{res.}\lesssim M_{pl}^{5}/(f\phi)^{5/2}$, to be evaluated at $\phi=\phi_{60}\sim 10 M_{pl}$ for the observable CMB scales. The bound thus finally reads $f_{NL}^{res.}\lesssim 3\times 10^{-3}\, (M_{pl}/f)^{5/2}$. The typical range for the axion decay constant is $10^{-4}M_{pl}\lesssim f\lesssim 0.1M_{pl}$ (see e.g.~\cite{Flauger:2009ab}). Consequently, for $f\gtrsim 5\times 10^{-2} M_{pl}$ this fundamental upper bound on $f_{NL}^{res.}$ becomes stronger, $f_{NL}^{res.}\lesssim {\cal O}(1)$ for $f\gtrsim 5\times 10^{-2} M_{pl}$, than the current observational bounds~\cite{Ade:2015ava}.

Secondly, we observe that in a quadratic potential the boundary to slow-roll eternal inflation (defined as the value of $\phi=\phi_\star$ where $\epsilon\sim V$) $\phi_\star\sim M_{pl}^{3/2}m^{-1/2}$ can be higher than our magnetic WGC field range bound $\phi<m^{-2/3}f^{1/3}M_{pl}^{4/3}$ for values of $f\lesssim 10^{-3} M_{pl}$, because COBE normalization of the CMB fluctuations fixes $m\sim 10^{-5}M_{pl}$. Intriguingly, recent analyses such as~\cite{Flauger:2014ana,Ade:2015lrj} (see also e.g.~\cite{Peiris:2013opa} for earlier work on the WMAP 9-year data) of the PLANCK data searching for oscillating contributions to the CMB power spectrum and the 3-point-function hint with the highest significance at very-high-frequency oscillating patterns with $f\sim 10^{-4}M_{pl}$. If this were corroborated in the future, then jointly with the magnetic WGC this would rule out slow-roll eternal inflation in quadratic axion monodromy inflation potentials in the past of our part of the universe.

We leave the generalization of both of these observations to more general axion monodromy potentials $V\sim \phi^p$ with `wiggles' as an interesting problem for the future.

\end{section}

\section*{Acknowledgements}

We thank J. J\" ackel, P. Mangat, E. Palti and L. Witkowski for useful discussions. This work was partly supported by the DFG Transregional Collaborative Research Centre TRR 33 ``The  Dark  Universe". F.R. is partially supported by the DAAD and the DFG Graduiertenkolleg GRK 1940 ``Physics Beyond the Standard Model''. The work of A.W. is supported by the ERC Consolidator Grant STRINGFLATION under the HORIZON 2020 contract no. 647995.


\begin{thebibliography}{nn}
{\small

\bibitem{Graham:2015cka} 
  P.~W.~Graham, D.~E.~Kaplan and S.~Rajendran,
  ``Cosmological Relaxation of the Electroweak Scale,''
  Phys.\ Rev.\ Lett.\  {\bf 115}, no. 22, 221801 (2015)
  doi:10.1103/PhysRevLett.115.221801
  [arXiv:1504.07551 [hep-ph]].


\bibitem{Espinosa:2015eda} 
  J.~R.~Espinosa, C.~Grojean, G.~Panico, A.~Pomarol, O.~Pujolas and G.~Servant,
  ``Cosmological Higgs-Axion Interplay for a Naturally Small Electroweak Scale,''
  arXiv:1506.09217 [hep-ph].


\bibitem{Hardy:2015laa} 
  E.~Hardy,
  ``Electroweak relaxation from finite temperature,''
  JHEP {\bf 1511}, 077 (2015)
  doi:10.1007/JHEP11(2015)077
  [arXiv:1507.07525 [hep-ph]].


\bibitem{Patil:2015oxa} 
  S.~P.~Patil and P.~Schwaller,
  ``Relaxing the Electroweak Scale: the Role of Broken dS Symmetry,''
  arXiv:1507.08649 [hep-ph].


\bibitem{Jaeckel:2015txa} 
  J.~Jaeckel, V.~M.~Mehta and L.~T.~Witkowski,
  ``Musings on cosmological relaxation and the hierarchy problem,''
  arXiv:1508.03321 [hep-ph].


\bibitem{Gupta:2015uea} 
  R.~S.~Gupta, Z.~Komargodski, G.~Perez and L.~Ubaldi,
  ``Is the Relaxion an Axion?,''
  arXiv:1509.00047 [hep-ph].


\bibitem{Batell:2015fma} 
  B.~Batell, G.~F.~Giudice and M.~McCullough,
  ``Natural Heavy Supersymmetry,''
  arXiv:1509.00834 [hep-ph].
  
\bibitem{Choi:2015fiu}
  K.~Choi and S.~H.~Im,
  arXiv:1511.00132 [hep-ph].

\bibitem{Kaplan:2015fuy} 
  D.~E.~Kaplan and R.~Rattazzi,
  ``A Clockwork Axion,''
  arXiv:1511.01827 [hep-ph].


\bibitem{Banks:2003sx} 
  T.~Banks, M.~Dine, P.~J.~Fox and E.~Gorbatov,
  ``On the possibility of large axion decay constants,''
  JCAP {\bf 0306}, 001 (2003)
  doi:10.1088/1475-7516/2003/06/001
  [hep-th/0303252].
  
\bibitem{Conlon:2016aea}
  J.~P.~Conlon and S.~Krippendorf,
  arXiv:1601.00647 [hep-th].
  
\bibitem{Abel:2015rkm}
  S.~Abel and R.~J.~Stewart,
  arXiv:1511.02880 [hep-th].


\bibitem{Kim:2004rp} 
  J.~E.~Kim, H.~P.~Nilles and M.~Peloso,
  ``Completing natural inflation,''
  JCAP {\bf 0501}, 005 (2005)
  doi:10.1088/1475-7516/2005/01/005
  [hep-ph/0409138].


\bibitem{Dimopoulos:2005ac} 
  S.~Dimopoulos, S.~Kachru, J.~McGreevy and J.~G.~Wacker,
  ``N-flation,''
  JCAP {\bf 0808}, 003 (2008)
  doi:10.1088/1475-7516/2008/08/003
  [hep-th/0507205].
  
\bibitem{Liddle:1998jc}
  A.~R.~Liddle, A.~Mazumdar and F.~E.~Schunck,
  Phys.\ Rev.\ D {\bf 58} (1998) 061301
  doi:10.1103/PhysRevD.58.061301
  [astro-ph/9804177].


\bibitem{Cicoli:2014sva} 
  M.~Cicoli, K.~Dutta and A.~Maharana,
  ``N-flation with Hierarchically Light Axions in String Compactifications,''
  JCAP {\bf 1408}, 012 (2014)
  doi:10.1088/1475-7516/2014/08/012
  [arXiv:1401.2579 [hep-th]].


\bibitem{Palti:2014kza} 
  E.~Palti and T.~Weigand,
  ``Towards large r from [p, q]-inflation,''
  JHEP {\bf 1404}, 155 (2014)
  doi:10.1007/JHEP04(2014)155
  [arXiv:1403.7507 [hep-th]].


\bibitem{Grimm:2014vva} 
  T.~W.~Grimm,
  ``Axion Inflation in F-theory,''
  Phys.\ Lett.\ B {\bf 739}, 201 (2014)
  doi:10.1016/j.physletb.2014.10.043
  [arXiv:1404.4268 [hep-th]].


\bibitem{Ibanez:2014kia} 
  L.~E.~Ibanez and I.~Valenzuela,
  ``The inflaton as an MSSM Higgs and open string modulus monodromy inflation,''
  Phys.\ Lett.\ B {\bf 736}, 226 (2014)
  doi:10.1016/j.physletb.2014.07.020
  [arXiv:1404.5235 [hep-th]].


\bibitem{Kappl:2014lra} 
  R.~Kappl, S.~Krippendorf and H.~P.~Nilles,
  ``Aligned Natural Inflation: Monodromies of two Axions,''
  Phys.\ Lett.\ B {\bf 737}, 124 (2014)
  doi:10.1016/j.physletb.2014.08.045
  [arXiv:1404.7127 [hep-th]].


\bibitem{Ben-Dayan:2014zsa} 
  I.~Ben-Dayan, F.~G.~Pedro and A.~Westphal,
  ``Hierarchical Axion Inflation,''
  Phys.\ Rev.\ Lett.\  {\bf 113}, 261301 (2014)
  doi:10.1103/PhysRevLett.113.261301
  [arXiv:1404.7773 [hep-th]].


\bibitem{Long:2014dta} 
  C.~Long, L.~McAllister and P.~McGuirk,
  ``Aligned Natural Inflation in String Theory,''
  Phys.\ Rev.\ D {\bf 90}, 023501 (2014)
  doi:10.1103/PhysRevD.90.023501
  [arXiv:1404.7852 [hep-th]].


\bibitem{Ben-Dayan:2014lca} 
  I.~Ben-Dayan, F.~G.~Pedro and A.~Westphal,
  ``Towards Natural Inflation in String Theory,''
  Phys.\ Rev.\ D {\bf 92}, no. 2, 023515 (2015)
  doi:10.1103/PhysRevD.92.023515
  [arXiv:1407.2562 [hep-th]].


\bibitem{Abe:2014xja} 
  H.~Abe, T.~Kobayashi and H.~Otsuka,
  ``Natural inflation with and without modulations in type IIB string theory,''
  JHEP {\bf 1504}, 160 (2015)
  doi:10.1007/JHEP04(2015)160
  [arXiv:1411.4768 [hep-th]].
 
\bibitem{Ruehle:2015afa}
  F.~Ruehle and C.~Wieck,
  JHEP {\bf 1505} (2015) 112
  doi:10.1007/JHEP05(2015)112
  [arXiv:1503.07183 [hep-th]].


\bibitem{Silverstein:2008sg} 
  E.~Silverstein and A.~Westphal,
  ``Monodromy in the CMB: Gravity Waves and String Inflation,''
  Phys.\ Rev.\ D {\bf 78}, 106003 (2008)
  doi:10.1103/PhysRevD.78.106003
  [arXiv:0803.3085 [hep-th]].


\bibitem{Marchesano:2014mla} 
  F.~Marchesano, G.~Shiu and A.~M.~Uranga,
  ``F-term Axion Monodromy Inflation,''
  JHEP {\bf 1409}, 184 (2014)
  doi:10.1007/JHEP09(2014)184
  [arXiv:1404.3040 [hep-th]].


\bibitem{Blumenhagen:2014gta} 
  R.~Blumenhagen and E.~Plauschinn,
  ``Towards Universal Axion Inflation and Reheating in String Theory,''
  Phys.\ Lett.\ B {\bf 736}, 482 (2014)
  doi:10.1016/j.physletb.2014.08.007
  [arXiv:1404.3542 [hep-th]].


\bibitem{Hebecker:2014eua} 
  A.~Hebecker, S.~C.~Kraus and L.~T.~Witkowski,
  ``D7-Brane Chaotic Inflation,''
  Phys.\ Lett.\ B {\bf 737}, 16 (2014)
  doi:10.1016/j.physletb.2014.08.028
  [arXiv:1404.3711 [hep-th]].


\bibitem{Blumenhagen:2014nba} 
  R.~Blumenhagen, D.~Herschmann and E.~Plauschinn,
  ``The Challenge of Realizing F-term Axion Monodromy Inflation in String Theory,''
  JHEP {\bf 1501}, 007 (2015)
  doi:10.1007/JHEP01(2015)007
  [arXiv:1409.7075 [hep-th]].


\bibitem{Hebecker:2014kva} 
  A.~Hebecker, P.~Mangat, F.~Rompineve and L.~T.~Witkowski,
  ``Tuning and Backreaction in F-term Axion Monodromy Inflation,''
  Nucl.\ Phys.\ B {\bf 894}, 456 (2015)
  doi:10.1016/j.nuclphysb.2015.03.015
  [arXiv:1411.2032 [hep-th]].


\bibitem{Ibanez:2014swa} 
  L.~E.~Ibanez, F.~Marchesano and I.~Valenzuela,
  ``Higgs-otic Inflation and String Theory,''
  JHEP {\bf 1501}, 128 (2015)
  doi:10.1007/JHEP01(2015)128
  [arXiv:1411.5380 [hep-th]].
  
\bibitem{Escobar:2015fda} 
  D.~Escobar, A.~Landete, F.~Marchesano and D.~Regalado,
  arXiv:1505.07871 [hep-th].
  
\bibitem{Kobayashi:2015aaa} 
  T.~Kobayashi, A.~Oikawa and H.~Otsuka,
  ``New potentials for string axion inflation,''
  arXiv:1510.08768 [hep-ph].
  
\bibitem{Escobar:2015ckf} 
  D.~Escobar, A.~Landete, F.~Marchesano and D.~Regalado,
  arXiv:1511.08820 [hep-th].
  



\bibitem{ArkaniHamed:2006dz} 
  N.~Arkani-Hamed, L.~Motl, A.~Nicolis and C.~Vafa,
  ``The String landscape, black holes and gravity as the weakest force,''
  JHEP {\bf 0706}, 060 (2007)
  doi:10.1088/1126-6708/2007/06/060
  [hep-th/0601001].


\bibitem{Conlon:2012tz} 
  J.~P.~Conlon,
  ``Quantum Gravity Constraints on Inflation,''
  JCAP {\bf 1209}, 019 (2012)
  doi:10.1088/1475-7516/2012/09/019
  [arXiv:1203.5476 [hep-th]].


\bibitem{Cheung:2014vva} 
  C.~Cheung and G.~N.~Remmen,
  ``Naturalness and the Weak Gravity Conjecture,''
  Phys.\ Rev.\ Lett.\  {\bf 113}, 051601 (2014)
  doi:10.1103/PhysRevLett.113.051601
  [arXiv:1402.2287 [hep-ph]].


\bibitem{Rudelius:2014wla} 
  T.~Rudelius,
  ``On the Possibility of Large Axion Moduli Spaces,''
  JCAP {\bf 1504}, no. 04, 049 (2015)
  doi:10.1088/1475-7516/2015/04/049
  [arXiv:1409.5793 [hep-th]].


\bibitem{delaFuente:2014aca} 
  A.~de la Fuente, P.~Saraswat and R.~Sundrum,
  ``Natural Inflation and Quantum Gravity,''
  Phys.\ Rev.\ Lett.\  {\bf 114}, no. 15, 151303 (2015)
  doi:10.1103/PhysRevLett.114.151303
  [arXiv:1412.3457 [hep-th]].


\bibitem{Rudelius:2015xta} 
  T.~Rudelius,
  ``Constraints on Axion Inflation from the Weak Gravity Conjecture,''
  JCAP {\bf 1509}, no. 09, 020 (2015)
  doi:10.1088/1475-7516/2015/09/020, 10.1088/1475-7516/2015/9/020
  [arXiv:1503.00795 [hep-th]].


\bibitem{Montero:2015ofa} 
  M.~Montero, A.~M.~Uranga and I.~Valenzuela,
  ``Transplanckian axions!?,''
  JHEP {\bf 1508}, 032 (2015)
  doi:10.1007/JHEP08(2015)032
  [arXiv:1503.03886 [hep-th]].


\bibitem{Brown:2015iha} 
  J.~Brown, W.~Cottrell, G.~Shiu and P.~Soler,
  ``Fencing in the Swampland: Quantum Gravity Constraints on Large Field Inflation,''
  JHEP {\bf 1510}, 023 (2015)
  doi:10.1007/JHEP10(2015)023
  [arXiv:1503.04783 [hep-th]].


\bibitem{Bachlechner:2015qja} 
  T.~C.~Bachlechner, C.~Long and L.~McAllister,
  ``Planckian Axions and the Weak Gravity Conjecture,''
  arXiv:1503.07853 [hep-th].


\bibitem{Hebecker:2015rya} 
  A.~Hebecker, P.~Mangat, F.~Rompineve and L.~T.~Witkowski,
  ``Winding out of the Swamp: Evading the Weak Gravity Conjecture with F-term Winding Inflation?,''
  Phys.\ Lett.\ B {\bf 748}, 455 (2015)
  doi:10.1016/j.physletb.2015.07.026
  [arXiv:1503.07912 [hep-th]].


\bibitem{Brown:2015lia} 
  J.~Brown, W.~Cottrell, G.~Shiu and P.~Soler,
  ``On Axionic Field Ranges, Loopholes and the Weak Gravity Conjecture,''
  arXiv:1504.00659 [hep-th].


\bibitem{Junghans:2015hba} 
  D.~Junghans,
  ``Large-Field Inflation with Multiple Axions and the Weak Gravity Conjecture,''
  arXiv:1504.03566 [hep-th].


\bibitem{Heidenreich:2015wga} 
  B.~Heidenreich, M.~Reece and T.~Rudelius,
  ``Weak Gravity Strongly Constrains Large-Field Axion Inflation,''
  arXiv:1506.03447 [hep-th].


\bibitem{Palti:2015xra} 
  E.~Palti,
  ``On Natural Inflation and Moduli Stabilisation in String Theory,''
  JHEP {\bf 1510}, 188 (2015)
  doi:10.1007/JHEP10(2015)188
  [arXiv:1508.00009 [hep-th]].


\bibitem{Heidenreich:2015nta} 
  B.~Heidenreich, M.~Reece and T.~Rudelius,
  ``Sharpening the Weak Gravity Conjecture with Dimensional Reduction,''
  arXiv:1509.06374 [hep-th].


\bibitem{Kooner:2015rza} 
  K.~Kooner, S.~Parameswaran and I.~Zavala,
  ``Warping the Weak Gravity Conjecture,''
  arXiv:1509.07049 [hep-th].
  
\bibitem{Andriot:2015aza}
  D.~Andriot,
  arXiv:1510.02005 [hep-th].

\bibitem{Kaloper:2015jcz} 
  N.~Kaloper, M.~Kleban, A.~Lawrence and M.~S.~Sloth,
  ``Large Field Inflation and Gravitational Entropy,''
  arXiv:1511.05119 [hep-th].


\bibitem{Ibanez:2015fcv} 
  L.~E.~Ibanez, M.~Montero, A.~Uranga and I.~Valenzuela,
  ``Relaxion Monodromy and the Weak Gravity Conjecture,''
  arXiv:1512.00025 [hep-th].


\bibitem{Kaloper:2008fb} 
  N.~Kaloper and L.~Sorbo,
  ``A Natural Framework for Chaotic Inflation,''
  Phys.\ Rev.\ Lett.\  {\bf 102}, 121301 (2009)
  doi:10.1103/PhysRevLett.102.121301
  [arXiv:0811.1989 [hep-th]].


\bibitem{Kaloper:2011jz} 
  N.~Kaloper, A.~Lawrence and L.~Sorbo,
  ``An Ignoble Approach to Large Field Inflation,''
  JCAP {\bf 1103}, 023 (2011)
  doi:10.1088/1475-7516/2011/03/023
  [arXiv:1101.0026 [hep-th]].
  
\bibitem{Dvali:2005an}
  G.~Dvali,
  hep-th/0507215.


\bibitem{Mazumdar:2014qea} 
  A.~Mazumdar and P.~Shukla,
  ``Model independent bounds on tensor modes and stringy parameters from CMB,''
  arXiv:1411.4636 [hep-th].


\bibitem{Chen:2008wn} 
  X.~Chen, R.~Easther and E.~A.~Lim,
  ``Generation and Characterization of Large Non-Gaussianities in Single Field Inflation,''
  JCAP {\bf 0804}, 010 (2008)
  doi:10.1088/1475-7516/2008/04/010
  [arXiv:0801.3295 [astro-ph]].


\bibitem{Flauger:2009ab} 
  R.~Flauger, L.~McAllister, E.~Pajer, A.~Westphal and G.~Xu,
  ``Oscillations in the CMB from Axion Monodromy Inflation,''
  JCAP {\bf 1006}, 009 (2010)
  doi:10.1088/1475-7516/2010/06/009
  [arXiv:0907.2916 [hep-th]].


\bibitem{Baumann:2006th} 
  D.~Baumann, A.~Dymarsky, I.~R.~Klebanov, J.~M.~Maldacena, L.~P.~McAllister and A.~Murugan,
  ``On D3-brane Potentials in Compactifications with Fluxes and Wrapped D-branes,''
  JHEP {\bf 0611}, 031 (2006)
  doi:10.1088/1126-6708/2006/11/031
  [hep-th/0607050].


\bibitem{Flauger:2014ana} 
  R.~Flauger, L.~McAllister, E.~Silverstein and A.~Westphal,
  ``Drifting Oscillations in Axion Monodromy,''
  arXiv:1412.1814 [hep-th].


\bibitem{Brown:1987dd} 
  J.~D.~Brown and C.~Teitelboim,
  ``Dynamical Neutralization of the Cosmological Constant,''
  Phys.\ Lett.\ B {\bf 195}, 177 (1987).
  doi:10.1016/0370-2693(87)91190-7


\bibitem{Brown:1988kg} 
  J.~D.~Brown and C.~Teitelboim,
  ``Neutralization of the Cosmological Constant by Membrane Creation,''
  Nucl.\ Phys.\ B {\bf 297}, 787 (1988).
  doi:10.1016/0550-3213(88)90559-7


\bibitem{Bousso:2000xa} 
  R.~Bousso and J.~Polchinski,
  ``Quantization of four form fluxes and dynamical neutralization of the cosmological constant,''
  JHEP {\bf 0006}, 006 (2000)
  doi:10.1088/1126-6708/2000/06/006
  [hep-th/0004134].


\bibitem{D'Amico:2012ji} 
  G.~D'Amico, R.~Gobbetti, M.~Kleban and M.~Schillo,
  ``Unwinding Inflation,''
  JCAP {\bf 1303}, 004 (2013)
  doi:10.1088/1475-7516/2013/03/004
  [arXiv:1211.4589 [hep-th]].


\bibitem{Coleman}
 S.~Coleman,
 \emph{The Use of Instantons} in \emph{Aspects of Symmetry, Selected Erice Lectures}, CUP 1985.

\bibitem{Coleman:1980aw} 
  S.~R.~Coleman and F.~De Luccia,
  ``Gravitational Effects on and of Vacuum Decay,''
  Phys.\ Rev.\ D {\bf 21}, 3305 (1980).
  doi:10.1103/PhysRevD.21.3305


\bibitem{Polchinski}
J.~Polchinski,
\emph{String Theory}, 2 vols., Cambridge University Press, 1998.

\bibitem{Giddings:2001yu}
  S.~B.~Giddings, S.~Kachru and J.~Polchinski,
  Phys.\ Rev.\ D {\bf 66} (2002) 106006
  doi:10.1103/PhysRevD.66.106006
  [hep-th/0105097].

\bibitem{Ade:2015ava} 
  P.~A.~R.~Ade {\it et al.} [Planck Collaboration],
  ``Planck 2015 results. XVII. Constraints on primordial non-Gaussianity,''
  arXiv:1502.01592 [astro-ph.CO].


\bibitem{Ade:2015lrj} 
  P.~A.~R.~Ade {\it et al.} [Planck Collaboration],
  ``Planck 2015 results. XX. Constraints on inflation,''
  arXiv:1502.02114 [astro-ph.CO].


\bibitem{Peiris:2013opa} 
  H.~Peiris, R.~Easther and R.~Flauger,
  ``Constraining Monodromy Inflation,''
  JCAP {\bf 1309}, 018 (2013)
  doi:10.1088/1475-7516/2013/09/018
  [arXiv:1303.2616 [astro-ph.CO]].
  
  

}


\end{thebibliography}
\end{document}